\documentclass[11pt]{article}
\pdfoutput=1
\usepackage{jcapmod}
\usepackage{booktabs}
\usepackage[english]{babel}
\usepackage{amsmath,amssymb,amsbsy,amstext, amsthm, simplewick}
\usepackage{graphicx}
\usepackage{amsfonts}
\usepackage{amssymb}
\usepackage{upgreek}
 \usepackage{exscale,relsize}
 \usepackage[makeroom]{cancel}
\usepackage{soul}
\usepackage{slashed}

\RequirePackage{color}

\usepackage{colortbl}
\definecolor{rp}{cmyk}{0.2, 1, 0.6, 0}
\definecolor{green2}{cmyk}{0, 1, 0.5, 0}
\definecolor{lightgreen}{cmyk}{0.2, 0, 0.2, 0.2}
\definecolor{lightgray}{cmyk}{0.1,0.2,0,0.1}
\definecolor{lightgray2}{cmyk}{0.4,0.4,0,0.8}
\definecolor{black}{cmyk}{1.0,1.0,1.0,1.0}

\allowdisplaybreaks[1]

\usepackage{colortbl}
\definecolor{lightgreen}{cmyk}{0.2, 0, 0.2, 0.2}
\definecolor{lightgray}{cmyk}{0.1,0.2,0,0.1}
\definecolor{lightgray2}{cmyk}{0.1,0.1,0,0.1}

\makeatletter
\newlength{\apb@width}
\newcommand{\autoparbox}[2][c]{\settowidth{\apb@width}{#2}\parbox[#1]{\apb@width}{#2}}

\makeatother

\numberwithin{equation}{section}

\def\beq{\begin{equation}}
\def\eeq{\end{equation}}

\def\bea{\begin{eqnarray}}
\def\eea{\end{eqnarray}}

\def\beq{\begin{equation}}
\def\eeq{\end{equation}}
\def\bea{\begin{eqnarray}}
\def\eea{\end{eqnarray}}

\def\H{{\cal H}}

\def\0{{\boldsymbol 0}}

\def\x{{\boldsymbol{x}}}

\def\H{{\cal H}}

\def\fnl{f_{\mathsmaller{\rm NL}}}
\def\fnlloc{f_{\mathsmaller{\rm NL}}^{\mathsmaller{\rm local}}}
\def\hatfnl{{\hat f}_{\mathsmaller{\rm NL}}}

\def\bng{\Delta b_{\mathsmaller{\rm NG}}}

\DeclareRobustCommand{\SkipTocEntry}[4]{}

\begin{document}

\begin{titlepage}

\setcounter{page}{1} \baselineskip=15.5pt \thispagestyle{empty}

\bigskip\

\vspace{1cm}
\begin{center}

{\fontsize{20}{28}\selectfont  \sffamily \bfseries Is There Scale-Dependent Bias \vspace{0.35cm}  

in Single-Field Inflation?}


\end{center}

\vspace{0.2cm}
\begin{center}
{\fontsize{13}{30}\selectfont  Roland de Putter,$^{\bigstar}$ Olivier Dor\'e,$^{\bigstar}$ and Daniel Green$^{\clubsuit}$}
\end{center}

\begin{center}

\vskip 8pt
\textsl{$^\bigstar$ Jet Propulsion Laboratory, California Institute of Technology, Pasadena, CA 91109 \\
\vskip 5pt
California Institute of Technology, Pasadena, CA 91125}
\vskip 7pt

\textsl{$^\clubsuit$ Canadian Institute for Theoretical Astrophysics, Toronto, ON M5S 3H8, Canada \\
\vskip 5pt
Canadian Institute for Advanced Research, Toronto, ON M5G 1Z8, Canada}\\

\end{center}

\vspace{1.2cm}
\hrule \vspace{0.3cm}
\noindent {Scale-dependent halo bias due to local primordial non-Gaussianity
provides a strong test of single-field inflation.
While it is universally understood that single-field inflation predicts negligible scale-dependent bias
compared to current observational uncertainties, there is still disagreement on the exact level of
scale-dependent bias at a level that could strongly impact inferences made from future surveys.
In this paper, we clarify this confusion and derive in various ways that
there is exactly zero scale-dependent bias in single-field inflation.
Much of the current confusion follows from the fact that single-field inflation
does predict a mode coupling of matter perturbations at the level of $\fnlloc \approx -5/3$,
which naively would lead to scale-dependent bias. However, we show explicitly that this mode coupling cancels out when perturbations are evaluated at a fixed physical scale rather than fixed coordinate scale.  Furthermore, we show how the absence of scale-dependent bias can be derived easily in any gauge.   This result can then be incorporated into a complete description of the observed galaxy clustering,
including the previously studied general relativistic terms, which are important
at the same level as scale-dependent bias of order $\fnlloc \sim 1$.
This description will allow us to draw unbiased conclusions about inflation from future galaxy clustering
data.

\vspace{0.3cm}
}

\hrule
\vskip 10pt

\vspace{0.6cm}
 \end{titlepage}

 \tableofcontents

\newpage

\section{Introduction}

Primordial non-Gaussianity is a powerful probe of the physics of inflation.
In particular, local type non-Gaussianity
can be used to discriminate
between single-field and multi-field models of inflation.
Single-field models generally predict negligible local non-Gaussianity \cite{Maldacena:2002vr,Creminelli:2004yq},
$\fnlloc \sim (n_s - 1) \ll 1$, such that any detection of non-zero local non-Gaussianity
would rule out the single-field scenario.
On the other hand, multi-field models quite generically
predict $|\fnlloc| \gtrsim 1$ \cite{lythetal03,zal04,toronto14}, setting a natural target for
the precision of
primordial non-Gaussianity searches at $\sigma(\fnlloc) \lesssim 1$.

Currently, the best constraint on local non-Gaussianity comes from the cosmic microwave
background (CMB) temperature and polarization bispectra, $\fnlloc = 0.8 \pm 5.0$ (68 \% confidence level)
\cite{planck2015png}, thus showing no evidence for deviations from Gaussianity,
while being about an order of magnitude away from the order unity target precision.
Unfortunately, due to the limited number of modes accessible with the CMB, it will not be possible to
improve this error bar by more than a factor of two with CMB data only \cite{cmbpol09}.
Fortunately, local primordial non-Gaussianity also leads to a strong signature in
cosmological large-scale structure, in the form of a scale-dependent halo bias $\bng(k) \propto k^{-2}$
\cite{Dalal:2007cu}.
The signal-to-noise of this effect is thus
peaked on very large scales and therefore is safely distinguishable from
the non-linearities of structure formation.
While systematics in the large-scale galaxy clustering measurement present a challenge,
it has been shown that large-volume future galaxy surveys can in principle
use this scale-dependent bias to reach the target precision $\sigma(\fnlloc) \lesssim 1$
\cite{ferramachoetal14,cameraetal14,ferrsmith14,raccetal14,yamauchietal14,RdPOli14}.

Since an order unity precision measurement of $\fnlloc$ is thus feasible,
 testing single-field inflation will require an equally precise prediction for scale-dependent bias in that class
of models. However, currently, there is still disagreement in the literature
about whether the scale-dependent bias is identically zero,
or instead non-zero at the level of an effective non-Gaussianity of order unity.
The difference in these predictions is small compared to current observational
error bars, both amounting to the statement that single-field inflation predicts
negligible scale-dependent bias, but for future surveys it is essential to
settle this question. For example, if the theoretical prediction is off by order unity,
we might end up falsely interpreting a measurement consistent with single-field inflation
as evidence for multi-field inflation.

The main focus of this paper is to resolve the confusion about scale-dependent bias
in single-field inflation and to derive a consistent answer in multiple ways.
The main conclusion will be that there is exactly zero scale-dependent bias in these models
so that any future detection of scale-dependent bias would indeed rule out single-field inflation.

The confusion in the literature derives from the contrast between two types of arguments.
On the one hand, the single-field consistency conditions are equivalent to stating
that, locally, a large-scale curvature fluctuation has no physical effect on small-scale
physics and is simply equivalent to a coordinate transformation. This strongly suggests (correctly) that
there can be no modulation of the halo number density by the long mode and therefore no scale-dependent bias~\cite{cremetal11,tanura11,baldaufetal11,Pajer:2013ana,hornhuixiao15,Hinterbichler:2012nm,daietal15}.
On the other hand, matter perturbations in synchronous gauge, for example,
do see a mode coupling at the level of an equivalent $\fnlloc \sim -5/3$
\cite{bartmatrio05},
so that the small-scale variance {\it is} modulated by a large-scale mode,
thus suggesting that there {\it is} scale-dependent bias \cite{verdemat09,brunietal14a,brunietal14b,villaetal14,cameraetal15,cameraetal15b}.
We will resolve this paradox by noting that the latter
is a modulation of the variance of matter perturbations on a fixed {\it coordinate} scale,
but that the local halo number density can only depend on
the variance on a given {\it physical} scale.
We will then
explicitly show that when the variance on a fixed physical scale is considered,
the modulation cancels out, leaving no physical effect, and therefore no scale-dependent bias.

Moreover, we will explain how both the original order unity mode coupling and the later cancellation
can be easily understood directly in terms of the single-field consistency conditions.
Specifically, making use to the work
of~\cite{Weinberg:2003sw, Fitzpatrick:2009ci, Creminelli:2013mca}, it
is easy to implement the initial conditions from single-field
inflation in any gauge.  The long wavelength mode
can be introduced by a change of coordinates and therefore it is easy to
determine its effect on any local quantity knowing only how it transforms
under a diffeomorphism\footnote{The same results can be derived in the
  Newtonian limit by \cite{Peloso:2013zw, Kehagias:2013yd} using
Galilean invariance.}.
This basic observation
has been applied to a number of large scale structure observables by a
number of authors (see
e.g.~\cite{Peloso:2013spa,Creminelli:2013poa,Kehagias:2013rpa,Nishimichi:2014jna,Horn:2014rta,
  Ben-Dayan:2014hsa,Mirbabayi:2014gda, Horn:2015dra, Kehagias:2015tda,
  Nishimichi:2015kla, Dai:2015jaa}).   
We apply this logic to the small-scale variance of matter perturbations to elegantly
derive the mode coupling, but we will show that it can even be applied to the halo number density itself
to directly obtain the result that $\bng (k) \equiv 0$ without referring to the modulation
of the variance of matter perturbations.

In addition to a precise prediction for scale-dependent bias,
there are other subtleties in the modeling of galaxy clustering,
which, if not properly taken into account, could lead to misinterpretation
of the data. What we observe in a survey are galaxy overdensities based on
positions derived from observed redshifts and observed angular positions,
which is not the same as the galaxy overdensity in some arbitrary given gauge.
The difference includes redshift-space distortions, but also more subtle relativistic
effects that contribute at order unity at the Hubble scale and thus are comparable in
magnitude to the scale-dependent bias effect from $\fnlloc \sim 1$.
In addition, one has to be careful that the very definition of bias is gauge independent.
These issues have all been addressed previously \cite{Yoo:2009au,Jeong:2011as,Yoo:2011zc}. In this paper,
we build on these results, in combination with the single-field prediction $\bng(k) \equiv 0$,
to present a complete prescription for modeling the galaxy clustering that can be used to test
primordial non-Gaussianity in an unbiased fashion.

\vskip 4pt
The paper is outlined as follows. In Section~\ref{sec:recipe},
we will provide a recipe for the observed galaxy
overdensity in single-field inflation that includes the proper relativistic definition of
galaxy bias, the (absence of) single-field scale-dependent bias, and
relativistic projection effects.
This section serves as a summary of our main result and places it in the context of
a complete galaxy clustering description.
In Section~\ref{sec:bias},
we will derive the scale-dependent bias in single-field inflation from
several perspectives.
We will reproduce many results
that have been derived in the literate but we will show how they can
be made compatible, leading to a single, consistent answer.
We will conclude in Section~\ref{sec:disc}. 

The paper includes two appendices. In Appendix~\ref{app:gauge}, we
show how to apply the inflationary consistency conditions in both Newtonian and
synchronous gauge.  We explain the slight differences that appear from
each perspective.  In Appendix~\ref{app:pert}, we show explicitly that
the mode coupling determined by second order perturbation theory is
precisely the same as the one determined by the consistency
conditions.  In this sense, the results of Section~\ref{sec:bias} can
be derived by a brute force calculation. 

Finally, for the remainder of this article, we will drop the ``${\rm local}$'' superscript
on $\fnl$, as we will exclusively discuss local type non-Gaussianity.

\section{The Observed Galaxy Bias in Single-Field Inflation}
\label{sec:recipe}

We wish to unambiguously express, in the context of single-field inflation, the {\it observed} galaxy density perturbations
in terms of underlying (first-order) cosmological perturbations that can be computed using Boltzmann codes
such as CAMB.
This problem can be divided into three components. First of all, even if we are given the fluctuations
in physical galaxy number density in some gauge,
$\delta_g$,
we need to transform this to the observed galaxy density fluctuations, $\Delta_g$,
which is estimated based on observed redshifts $z$ and angular positions ${\bf \hat{n}}$.
Secondly, to obtain $\delta_g$, we need a way to define galaxy bias that does not suffer from gauge ambiguities.
Finally, we need to specify the scale-dependence (if any) of the thus defined galaxy bias parameter.

The main focus of this paper is to compute and clarify the third
component, i.e.,~is there a scale-dependent galaxy bias in
single-field inflation? We will discuss this in detail in Section \ref{sec:bias}.
In the current section, we simply use the main result of Section \ref{sec:bias}
as part of a complete recipe for the observed galaxy clustering
in the context of single-field inflation, containing all three of the components discussed above.

First, the relation between observed galaxy overdensity and
the density and metric perturbations in a given gauge
has been addressed definitively in the literature, e.g.~\cite{Yoo:2009au,yoo10,challlew11,brunietal12,Jeong:2011as,bonvindurrer11,Yoo:2011zc, bertetal12}.
For instance, in terms of synchronous gauge fluctuations, where (in a spatially flat universe)
\bea
\label{eq:synch}
ds^2 &=& a^2(\tau) \left[ -  d\tau^2 + (1 - 2 \hat{\psi}) \, \delta_{ij}^{\rm K} \, dx^i dx^j +  \partial_i \partial_j \chi \, dx^i dx^j\right],
\eea
and where velocity perturbations in cold dark matter are set to zero,
\cite{Jeong:2011as} gives (borrowing here some notation from \cite{bertetal12})
\beq
\label{eq:obs dg gen}
\Delta_g({\bf \hat{n}}, z) = \Delta_{g,{\rm loc}}({\bf \hat{n}}, z) + \Delta_{g,\kappa}({\bf \hat{n}}, z) + \Delta_{g,I}({\bf \hat{n}}, z),
\eeq
with the ``local'' contribution,
\bea
\label{eq:obs dg}
\Delta_{g,{\rm loc}} &=& \delta_g + \frac{1}{2} \left[b_e - \left(1 + 2 \mathcal{Q} \right) + \frac{1+z}{H}\,\frac{dH}{dz}
-\frac{2}{D} \, \left( 1 - \mathcal{Q} \right) \, \frac{1+z}{H} \right] \left( \partial_\parallel \chi' + \chi''\right) \nonumber \\
&-& \frac{1+z}{2 H} \, \partial^2_\parallel \chi '
- \frac{2}{D}\,\left( 1 - \mathcal{Q}\right)\left( D \, \hat{\psi} + \frac{1}{2} \chi'\right),
\eea
and $\Delta_{g,\kappa}$ and $\Delta_{g,I}$ given in \cite{bertetal12} as line-of-sight integrals
over metric perturbations. We ignore the stochastic noise component here.
For this expression to be valid, it is crucial that the perturbations on the right hand side are evaluated
in synchronous gauge (of course, $\Delta_g$ can in principle be expressed in terms of perturbations in any gauge).
In Eq.~(\ref{eq:obs dg}), $\delta_g$ is the physical galaxy overdensity in synchronous gauge,
$b_e$ the evolution bias, $b_e = d\ln(a^3 \bar{n}_g)/d\ln a$ (with $\bar{n}_g$ the mean physical number density),
$\mathcal{Q}$ is the response to magnification $\delta \mathcal{M}$, defined such that magnification bias causes
$\Delta_g \to \Delta_g + \mathcal{Q} \, \delta \mathcal{M}$, and $D$ is comoving distance to the galaxy sample.

Second, we need to relate the galaxy overdensity $\delta_g$ to the underlying matter (and metric) perturbations by
a bias prescription. The standard approach is to define the galaxy bias $b$ as the proportionality factor between
$\delta_g$ and the matter overdensity $\delta$. However, this makes the definition of $b$ gauge-dependent.
The natural gauge choice is the synchronous gauge (e.g.~\cite{Jeong:2011as}), as here slices of constant coordinate time $\tau$
corresponds to constant proper time so that there is no evolution bias. Thus,
\beq
\label{eq:synch bias}
\delta_g = b \, \delta \quad {\rm \it (synchronous \, gauge)}
\eeq
defines the bias.
In an arbitrary gauge, physical density contrast is replaced by a ``gauge invariant" quantity
\beq\label{equ:ginvariant}
\delta_g^{\rm Invariant}\equiv \delta_g + \frac{d\ln\bar{n}_g}{d\ln a} \, a H \delta \tau = b \, \left( \delta - 3 a H \delta \tau \right) ,
\eeq
where $\delta \tau$ is the coordinate time shift between synchronous gauge and
the gauge under consideration\footnote{i.e.~$\tilde{\tau} = \tau + \delta \tau$,
with $\tau$ the conformal time coordinate in synchronous gauge and
$\tilde{\tau}$ the conformal time in the gauge under consideration}.
Note that the ratio $\delta_g/\delta$ is clearly gauge-dependent.
In particular, if the true bias $b$ as defined above is scale-independent, the ratio $\delta_g/\delta$
in a gauge where constant coordinate time slices do not correspond to constant proper time will
in general be scale-dependent. It is thus important that we have first defined $b$ in a gauge-independent
manner before we address scale-dependent bias.

The third component that needs to be addressed then is the specification of the galaxy bias $b$ defined in Eq.~(\ref{eq:synch bias}),
and specifically its scale-dependence. To linear order, and for Gaussian initial conditions,
the bias is scale-independent and can thus be described by a single free parameter,
$b = b_1$.  However, in the presence of primordial non-Gaussianity,
there is an additional, scale-dependent bias contribution. Specifically, if the non-Gaussianity is of local type, a large-scale potential fluctuation
will modulate the initial variance of matter fluctuations on small scales, thus modulating
the number density of halos that these fluctuations will collapse into at a later time.
Working in synchronous gauge, and going to Fourier space for convenience,
\beq
\label{eq:db}
\delta_{g,l} = b_1 \, \delta_l + \frac{d\ln n_g}{d\ln \sigma_R} \, \frac{d\ln \sigma_R}{d\varphi_{l}} \, \varphi_{l}
= \left(b_1 - \frac{3 H_0^2 \Omega_m}{2 T(k) D(z) k^2} \,  \frac{d\ln n_g}{d\ln \sigma_R} \, \frac{d\ln \sigma_R}{d\varphi_{l}} \right) \, \delta_l.
\eeq
Here, $\varphi$ is the metric perturbation evaluated
in the early universe\footnote{The description here in terms of $\varphi_l$ is
technically only correct if the initial conditions
are taken during matter domination with the long mode $\varphi_l$ outside of the horizon,
which is not possible of the mode entered the horizon during radiation domination.
In Section \ref{sec:bias}, we will use a more general description in terms of the curvature
perturbation $\zeta_l$, but here we use $\varphi_l$ since this is a very common description
in the literature.},
related to the matter overdensity $\delta$ at redshift $z$ by $k^2 \varphi = -3H_0^2 \Omega_m/2T(k)D(z)k^2$
(using the Poisson equation),
where $T(k)$ is the transfer function
of matter perturbations, normalized to $1$ at low $k$, $D(z)$ is the linear
growth function, normalized such that $D(z) = 1/(1 + z)$ during matter
domination, and $\sigma_R^2$ is the initial variance of
small-scale fluctuations smoothed over a comoving length $R$
enclosing the mass of the
halos of interest (see Eq.~\ref{eq:sigR}).
The quantity $n_g$
is the galaxy number density in a region with initial small-scale variance $\sigma^2_R$.
Technically, we have thus assumed here that galaxies occupy halos of a single mass.
More generally, the dependence on the variance $\sigma^2_R$
would be replaced by a dependence on the variance on a range of scales corresponding to a range of halo masses.
We have applied an explicit subscript $l$ to stress that we are looking at the effect of large-scale fluctuations
compared to the fluctuations that collapse into halos.

Thus, in the presence of mode coupling (non-zero $d\ln \sigma_R/d \varphi_l$), the bias gets a scale-dependent correction,
\beq
\label{eq:bias gen}
b(k) = b_1 +\bng(k) \ ,
\eeq
where $\bng(k) \propto k^{-2}$.
If we write the mode coupling in terms of an effective local non-Gaussianity parameter,
\beq
\frac{d\ln\sigma_R}{d\varphi_{l}} = - 2 \, \fnl^{(\Delta b)} \, ,
\eeq
we obtain for the scale-dependent bias\footnote{Note that if we assume a universal halo mass function, the response
function of the galaxy number density simply becomes,
\beq
\frac{d\ln n_g}{d\ln \sigma_R} = (b_1 - 1) \, \delta_c,
\eeq
so that Eq.~(\ref{eq:db gen}) reduces to the scale-dependent bias first discovered in \cite{Dalal:2007cu}.},
\beq
\label{eq:db gen}
\bng(k) = 2 \,  \fnl^{(\Delta b)} \, \frac{d\ln n_g}{d\ln \sigma_R} \, \frac{3 H_0^2 \Omega_m}{2 T(k) D(z) k^2}.
\eeq

There has been significant confusion about the level of scale-dependent
bias, i.e.~of the mode coupling characterized by $f_{\rm NL}^{(\Delta b)}$ in Eq.~(\ref{eq:db gen}), in single-field inflation. On the one hand, naively,
$\fnl^{(\Delta b)}$ is simply equal to the local non-Gaussianity parameter
of $\zeta$ (the comoving curvature perturbation on slices of constant density),
which in turn is small and given by the single-field consistency condition \cite{Maldacena:2002vr,Creminelli:2004yq},
suggesting that the scale-dependent
bias is
\beq
\fnl^{(\Delta b)}  \stackrel{?}{=} - \frac{5}{12} (n_s  - 1) \, ,
\eeq
where $n_s$ is the tilt of the primordial power spectrum.
On further inspection, since $\zeta$ and $\delta$ are non-linearly related, their non-Gaussianity parameters are not the same.
In synchronous gauge, it has been found that the non-Gaussianity of matter perturbations is described by an effective non-Gaussianity
parameter that gets shifted by $-5/3$ (e.g.~\cite{bartmatrio05}). This has led several authors to conclude that single-field inflation predicts
scale-dependent bias of order unity \cite{verdemat09,brunietal14a,brunietal14b,villaetal14} (see also \cite{cameraetal15,cameraetal15b}),
\beq
\fnl^{(\Delta b)}  \stackrel{?}{=} -\frac{5}{3} - \frac{5}{12} (n_s  - 1) \, .
\eeq
On the other hand, however, basic physical arguments suggest that, modulo gradients, the metric perturbation $\varphi_l$
cannot have any effect on small-scale physics and its effect is equivalent to a coordinate transformation.
This strongly suggests that the scale-dependent bias has to be zero
\cite{cremetal11,tanura11,baldaufetal11,Pajer:2013ana,hornhuixiao15,Hinterbichler:2012nm,daietal15}:
\beq
\fnl^{(\Delta b)}  \stackrel{?}{=} 0 \, .
\eeq

How are these result compatible? The main goal of this paper is to clarify this confusion.
In the next section, we will calculate $\bng$ both using the mode coupling of density fluctuations in synchronous gauge
and using the argument that $\varphi_l$ corresponds to a simple coordinate transformation, and will show that
both approaches lead to the same result and that the former approach can be understood easily in terms of the latter.
We aim to present a pedagogical presentation in order to settle the confusion as much as possible.
The goal of the current section is simply to present the end result, which is that {\it in single-field inflation,
there is no scale-dependent bias}:
\beq
\label{eq:zero ng bias}
\boxed{ \fnl^{(\Delta b)}  \equiv 0}\ .
\eeq

In general, the observed galaxy clustering is thus fully described by Eqs~(\ref{eq:obs dg gen})
and Eq.~(\ref{eq:obs dg}), 
with the galaxy overdensity given by
Eq.~(\ref{eq:synch bias}), and the bias by Eq.~(\ref{eq:bias gen}).
In other words, in synchronous gauge, the galaxy overdensity entering Eq.~(\ref{eq:obs dg}) is
$\delta_g({\bf k}) = (b_1 + \bng (k)) \, \delta({\bf k})$
so that in single-field inflation,
\beq
\boxed{ \bng(k) \equiv 0 \, .}
\eeq
Note however that the conversion from physical galaxy overdensity to observed galaxy overdensity in
Eq.~(\ref{eq:obs dg}) does introduce terms that have the same scale-dependence as the scale-dependent bias
would have. However, if after taking these well understood terms into account, there is remaining evidence for a true
scale-dependent bias contribution $\bng \propto k^{-2}$, this would rule out single-field inflation.  

While we have in this section expressed the observed galaxy overdensity in terms
of synchronous gauge variables, our definition of bias was gauge invariant (Eq.~(\ref{equ:ginvariant}))
so that
our result can be applied to obtain $\Delta_g$ starting from any gauge.  Existing calculations in Newtonian gauge can be written
in terms of the same bias definition (see e.g.~\cite{Yoo:2011zc})
and therefore no special care is needed and we may again use $\bng = 0$.

\section{Single-Field Consistency and Derivations of Scale-Dependent Bias}
\label{sec:bias}

In this section, we will provide several derivations of the result $\bng = 0$
for initial conditions satisfying the single-field consistency relation.
We will start by reviewing the inflationary single-field consistency condition itself in
Section \ref{subsec:maldacena}. 
We will then derive the matter density mode coupling between long and short wavelengths
using the same arguments that led to the inflationary consistency condition
in Section \ref{subsec:modecoupling}
and we confirm the result by explicitly matching the second order perturbation theory
solution of matter perturbations to the inflationary consistency relation (Appendix \ref{app:pert}).
While we consistently find the same non-zero mode coupling in synchronous gauge, for example, it does not
lead to scale-dependent bias because the modulation of short modes on a given {\it physical} scale
is zero. We
will show explicitly that the conversion from short modes on a fixed
coordinate scale to modes on a fixed physical scale
exactly cancels the aforementioned mode coupling and our derivation will make it clear that this cancellation is inevitable, as it is a direct manifestation
of the consistency condition.
We will finally see in Section \ref{subsec:direct} how this result can be derived more simply by applying the consistency conditions to the galaxies directly.
All together, we will provide three different perspectives that can be applied in any gauge to derive the same physical result.

\subsection{Single-field consistency conditions}
\label{subsec:maldacena}

In order to build intuition, let us first review the derivation of the consistency conditions for $\zeta$,
the comoving curvature perturbation on slices of constant energy density \cite{BST83,salopekbond90}.
If we work on slices of constant density, on super-horizon scales (i.e.~neglecting gradients
of the curvature perturbation), the metric takes the general form,
\beq
ds^2 = a^2(\tau) \left[ - d\tau^2 + e^{2 \zeta({\bf x})} \,  d{\bf x}^2 \right].
\eeq
Dividing the curvature perturbation into a large- and small-scale
contribution, $\zeta \equiv \zeta_s + \zeta_l$,
the single-field consistency
conditions are derived by observing that
a large-scale perturbation can,
to zeroth order in gradients of $\zeta_l$,
be generated by a diffeomorphism.
Specifically,
we can locally describe the perturbations in a coordinate system (denoted by a tilde) where there is no large-scale mode,
\beq
\tilde{ds^2} = a^2(\tau) \, e^{2 \tilde{\zeta}_s({\bf \tilde{x}})} \,  d{\bf \tilde{x}}^2,
\eeq
where $\tilde{\zeta}_s({\bf \tilde{x}})$, and more generally the local physics, knows nothing of the existence of a large-scale mode.
We have written only the space-space contribution to the line element
(as there are only spatial metric perturbations in this gauge),
and we do not apply a tilde to $\tau$ because we will not apply any transformation to it.
The large-scale mode is then generated by applying a diffeomorphism to a new coordinate system
(without a tilde), defined by the dilation,
\beq
\label{eq:coord rescaling}
{\bf x} \equiv {\bf \tilde{x}} \, (1 - \zeta_l) \, , 
\eeq
leaving the time coordinate unchanged.
When $\zeta_l$ is a constant, this is a
pure gauge mode.  However, to leading order in derivatives we may
modify $\zeta_l \to \zeta_l (\x)$ and this matches onto a physical
long wavelength solution up to higher derivative corrections.
Since the line element itself is a scalar,
in the new coordinate system, we then have, to first order in $\zeta_l$,
\beq
\label{eq:zeta trans}
ds^2 = \tilde{ds^2} = a^2(\tau) \, e^{2 \tilde{\zeta}_s({\bf \tilde{x}})} \,  d{\bf \tilde{x}}^2 
= a^2(\tau) \, e^{2 \tilde{\zeta}_s({\bf \tilde{x}})} \, (1 + 2 \zeta_l) \, d{\bf x}^2 
= a^2(\tau) \, e^{2 (\tilde{\zeta}_s({\bf \tilde{x}}) + \zeta_l)} \, d{\bf x}^2,
\eeq
where the final equality is true to first order\footnote{We could have generated
the large mode non-perturbatively with the dilation ${\bf x} \equiv {\bf \tilde{x}} \, e^{-\zeta_l}$,
but since the long mode is assumed to be small, we are content to work to first order in $\zeta_l$.} in $\zeta_l$.

Comparing the far left-hand side of Eq.~(\ref{eq:zeta trans}) to the far right hand-side
shows first of all that the coordinate transformation indeed generates the large-scale curvature perturbation $\zeta_l$,
and, secondly, that the small-scale curvature perturbation transforms as a scalar, i.e.~$\zeta_s({\bf x}) = \tilde{\zeta}_s({\bf \tilde{x}})$.
Hence, in this ${\bf x}$ coordinate system where the large-scale mode is manifest,
\beq
\label{eq:consistency simple}
\zeta_s({\bf x}) =  \tilde{\zeta}_s({\bf x} \, (1 + \zeta_l)) = \tilde{\zeta}_s({\bf x}) + \zeta_l \, {\bf x} \cdot \nabla \tilde{\zeta}_s({\bf x}),
\eeq
with $\tilde{\zeta}_s({\bf x})$ statistically independent of $\zeta_l$.
This is the {\it squeezed limit, single-field consistency relation.}
Note that, in formulating the mode coupling, it was crucial that the short mode on the right-hand side
is statistically independent of the long mode. This is equivalent to it being the small-scale component of a Gaussian field since
for Gaussian fields, different scales are independent. Thus we could also write $\tilde{\zeta}_s \equiv \zeta_s^G$.

We now wish to relate the mode coupling in Eq.~(\ref{eq:consistency simple}) to a non-Gaussianity parameter $f_{\rm NL}$.
The usual {\it ansatz} is
\beq
\zeta = \zeta^G + \frac{3}{5} \, \fnl \, \left( (\zeta^{G})^2 - \langle (\zeta^{G})^2 \rangle \right),
\eeq
where $\zeta^{G}$ is a Gaussian field. This leads to the mode coupling,
\beq
\label{eq:fnl long-short}
\zeta_s = \left( 1 + \frac{6}{5} \, \zeta_l \, \fnl \right) \, \zeta_s^G.
\eeq
Comparing to the consistency condition, Eq.~(\ref{eq:consistency simple}), implies that the mode coupling is really described
by a mode coupling {\it operator},
\beq
\hatfnl = \frac{5}{6} \, {\bf x} \cdot \nabla \, .
\eeq
However, in a statistical sense, we can replace this operator by an equivalent numerical value.  The operator ${\bf x} \cdot \nabla$ corresponds to a rescaling of coordinates so that, statistically, we expect
it to correspond to the logarithmic derivative of the root-mean-square fluctuation in $\zeta_s$, i.e.~
\beq
\fnl = - \frac{5}{6} \, \frac{d\ln \sqrt{\Delta^2_\zeta}(k)}{d\ln k} = -\frac{5}{12} \, (n_s - 1).
\eeq
(the minus sign follows from going to Fourier space) where $\Delta^2_\zeta(k) = \frac{k^3 P_\zeta(k)}{2 \pi^2}$ is the dimensionless power spectrum of $\zeta$.  This is how the consistency relation is commonly phrased.
A more rigorous derivation of $\fnl$ would involve explicitly calculating
the effect of
the mode coupling $\hatfnl$ on the squeezed limit bispectrum, or equivalently, on
the modulation of the small-scale variance by the long mode,
and equating this to the bispectrum predicted by the local ansatz in (\ref{eq:fnl long-short}).  Details can be found in many references, including~\cite{Maldacena:2002vr, Creminelli:2004yq}.

In the following, we will often make use of both forms for the non-Gaussianity,
i.e.~as an operator and as a number,
but we emphasize that the latter description is only valid in a statistical sense and not
at the level of Eq.~(\ref{eq:fnl long-short}).
Note also that the statistical equivalence of ${\bf x} \cdot \nabla$ to $-d\ln \sqrt{\Delta^2}(k)/d\ln k$ is generally valid,
but care has to be taken to use the dimensionless power spectrum $\Delta^2(k)$ of the field on which the operator acts.
For instance, for the matter perturbations, we would get ${\bf x} \cdot \nabla \to -1/2 \, (3 + n_s)$. This will play an important role in the next subsection. 
\vskip 6pt
The key insight to be learned from this discussion is that, up to derivatives\footnote{The {\it conformal consistency conditions} would allow us to extend this procedure to include one gradient of the long mode~\cite{Creminelli:2012ed,Hinterbichler:2012nm}.  We will not include them here in the interest of pedagogy.}  of
$\zeta_l$, we can understand the effect of the long mode by performing
a diffeomorphism. This insight applies in any gauge and to any
quantity of interest, provided one knows how it transforms. It is this
feature that is a prediction of single-field inflation which we will
call the {\it single-field consistency conditions}, rather than the specific value of $\fnl$.  

The single-field consistency conditions follow from the fact that during single-field inflation
there is a single ``clock'', see e.g.~\cite{Creminelli:2004yq}.
This assumes the evolution has reached the inflationary attractor solution (although see \cite{mooijetal15}
for a generalization to non-attractor solutions) and
ignores decaying modes.
The consistency conditions remain valid even after inflation, as long as perturbations evolve adiabatically (so that
there still is only a single clock) \cite{Weinberg:2003sw,Creminelli:2013mca}. Adiabaticity in turn is maintained as long as gravity obeys the equivalence
principle and as long as scales larger than the sound horizon are considered for the long mode. Thus,
since we assume general relativity,
we can always apply the consistency conditions while the long mode is super-horizon,
and after radiation domination we can even apply them over a large range of scales
inside the horizon.
We will make use of this late-universe application of the consistency conditions in the following section
to derive the mode coupling of matter perturbations and to derive scale-dependent galaxy/halo bias.

The application to the late universe has
recently been highlighted by~\cite{Creminelli:2013mca}
(based also on~\cite{Fitzpatrick:2009ci}), where they were applied directly to the matter
perturbations in Newtonian gauge.  The techniques used in large-scale structure follow from an observation
of Weinberg~\cite{Weinberg:2003sw} (which was generalized in~\cite{Hinterbichler:2012nm})
that, in the context of single-field inflation, a long wavelength adiabatic mode can be generated by a diffeomorphism that
leaves the gauge fixed.  We will follow the same logic, although we will specialize
to synchronous gauge (see Appendix~\ref{app:gauge} for more details on the relation to Newtonian gauge).

\subsection{Scale-dependent bias from mode coupling}
\label{subsec:modecoupling}

We now turn to the calculation of scale-dependent bias in single-field inflation using the
mode coupling between long and short modes. We again separate perturbations into long and short wavelength
components, $\delta = \delta_s + \delta_l$, etc.
We work in synchronous gauge, where the metric is given by Eq.~(\ref{eq:synch}).  In this gauge, constant $\tau$ slices are hypersurfaces
of constant proper time.
Using the fact that galaxies live in dark matter halos, we can then express the local number density of galaxies at some time in the late universe
in terms of the initial matter perturbations as\footnote{We are implicitly taking an excursion set formalism approach to halo formation
and we are considering galaxies living in halos of a fixed mass enclosed in an initial comoving radius $R$.
However, all our conclusions are valid in general as long as galaxies live in halos and halos correspond to overdensities
in the initial matter distribution.}
\beq
\label{eq:ng ansatz}
n_g = n_g(\delta_l, \sigma^2_{R}).
\eeq
In another gauge, different locations at constant coordinate time might be in different stages of their evolution so that
the galaxy number density would also explicitly depend on the proper time fluctuations. It is thus crucial that we chose a synchronous gauge
to start from (see e.g.~\cite{Jeong:2011as}).

Eq.~(\ref{eq:ng ansatz}) says, first of all, that $n_g$ has an explicit
dependence on the initial $\delta_l$. This is because the long mode affects the evolution of
the small-scale perturbations and therefore
the abundance of collapsed objects at a later time.
Secondly, there is an explicit dependence on $\sigma_R^2$,
the initial variance of small-scale fluctuations on a scale corresponding to the halos of interest.
In the presence of primordial non-Gaussianity, this variance can in principle
be modulated by the large-scale metric perturbation $\zeta_l$, leading to scale-dependent bias.
To quantify the scale-dependent bias in the single-field case, we thus need
to compute the modulation
\beq
\frac{d\sigma^2_{R}}{d\zeta_l} \ .
\eeq
We choose to describe the long mode in terms of the curvature perturbation $\zeta$ because it
is conserved outside of the horizon.
It is common to instead
write the modulation in terms of the initial Newtonian potential during matter domination, $\varphi_l = -3/5 \, \zeta_l$
(see also Appendix~\ref{app:pert}).
However, note that if the long-wavelength mode entered the horizon
before matter-radiation equality, it is not possible to consider (super-horizon) initial conditions during
matter domination.

The modulation gets two contributions:
\begin{itemize}
\item
There is a mode coupling between $\delta_s$ and the long mode,
\beq
\label{eq:define fnleff}
\delta_s =  \left(1 + \frac{6}{5} \, \zeta_l \, \hat{f}_{\rm NL}^{\rm eff}  \right) \, \delta_s^G \ ,
\eeq
where we have implicitly defined $\hat{f}_{\rm NL}^{\rm eff}$ analogously to the non-Gaussianity parameter for $\zeta$,
cf.~Eq.~(\ref{eq:fnl long-short}).
This gives a
modulation of the small-scale variance on a fixed coordinate scale $R$
of 
\beq
(\sigma^2_R)_{\varphi_l} = \left(1 + \frac{12}{5} \, \zeta_l \, f_{\rm NL}^{\rm eff} \right) \, \sigma^2_R \ .
\eeq
Note that $f_{\rm NL}^{\rm eff}$ will be different than the $f_{\rm NL}$ appearing in
Eq.~(\ref{eq:fnl long-short}) because the matter density has a nonlinear relation to $\zeta$.
The small-scale variance is
\beq
\label{eq:sigR}
\sigma_R^2 = \int d^3 {\bf k}\, |W(k R)|^2 \, P_\delta(k) \ .
\eeq
Here, $W(kR)$ is the Fourier transform of, e.g., a spherical top-hat function enclosing a sphere with
radius $R$ and $P_\delta(k)$ is the matter power spectrum.
\item
The coordinate scale on which the variance is computed should be
the radius $R$ corresponding to a fixed {\it physical} lenght scale, $a(\tau) \, \tilde{R}$.
This means that the coordinate scale $R$
is modulated by $\zeta_l$.
\end{itemize}
We will now show that the above two effects exactly cancel out in single-field inflation\footnote{In many cases, the above procedure is performed in Lagrangian coordinates.  In that case, it has been shown that the consistency conditions acts trivially~\cite{Horn:2015dra}.}.

In synchronous gauge, we can generate the long mode in the same way we did in comoving
gauge\footnote{The technical subtleties that have been discussed in comoving gauge and Newtonian gauge
for obtaining physical solutions are not important in synchronous gauge.
See Appendix~\ref{app:gauge} for details.}
to derive the inflationary consistency conditions in Section \ref{subsec:maldacena}.
We again start in the local coordinate frame where the
mode $\zeta_l$ is taken out. Here 
\beq
\label{eq:rho_sync}
\tilde{\rho}({\bf \tilde{x}},\tau) = \bar{\rho}(\tau) \, \left( 1 + \tilde{\delta}_s({\bf \tilde{x}}, \tau) \right) \, ,
\eeq
where the perturbation $\tilde{\delta}_s({\bf \tilde{x}}, \tau)$ is independent of the long mode.
Now we again apply the spatial coordinate rescaling, Eq.~(\ref{eq:coord rescaling}), that makes the long mode explicit.
Since we do not transform $\tau$, and the matter density is a scalar, we have
\beq
\bar{\rho}(\tau) \, \left(1 + \delta({\bf x}, \tau) \right) = \rho({\bf x}, \tau) = \tilde{\rho}({\bf \tilde{x}}, \tau)
=  \bar{\rho}(\tau) \, \left( 1 + \tilde{\delta}({\bf \tilde{x}}, \tau) \right) \, ,
\eeq
so that $\delta_s$ transforms as a scalar (as did $\zeta_s$) under the dilation,
\beq
\delta_s({\bf x},\tau) = \tilde{\delta}_s({\bf \tilde{x}},\tau) \, .
\eeq
Thus, in the coordinate system with the long mode explicit, the mode coupling takes the same form as for $\zeta$,
\beq
\label{eq:delta_s}
\delta_s({\bf x},\tau) = \tilde{\delta}_s({\bf x} (1 + \zeta_l)) = \tilde{\delta}_s({\bf x}) + \zeta_l \, {\bf x} \cdot \nabla \tilde{\delta}_s({\bf x}) \, ,
\eeq
or,
\beq
\label{eq:diff mode coupling}
\delta_s^{(2)} = 2 \, \zeta_l \, {\bf x} \cdot \nabla \tilde{\delta}_s  \, .
\eeq
Here we have expanded the matter perturbation $\delta = \delta^{(1)} + \frac{1}{2} \delta^{(2)} + \dots$.
An alternative way to derive this mode coupling, see e.g.~\cite{verdemat09,brunietal14a,brunietal14b,villaetal14},
is to compute the second order perturbation theory solution
for matter perturbations in the post-inflationary epoch and to match this solution to the consistency
condition for $\zeta$ given by single-field inflation. We follow this approach for the special case of a matter dominated
Universe in Appendix~\ref{app:pert} and show that it yields the exact same result found in a more straightforward way above.

Comparing this to the definition of the effective non-Gaussianity for the matter perturbation,
Eq.~(\ref{eq:define fnleff}),
\beq
\label{eq:delta_s2}
\delta_s^{(2)} = \frac{12}{5} \, \zeta_l \, \hat{f}_{\rm NL} \,  \delta_s^{(1)} \, ,
\eeq
we find $\hat{f}_{\rm NL}^{\rm eff} = \frac{5}{6} \, {\bf x} \cdot {\nabla}$.
Statistically, the operator ${\bf x} \cdot {\nabla}$
acting on $\delta_s^{(1)}$ gives $-\frac{d \sqrt{\Delta_\delta(k)}}{d\ln k} = - \frac{1}{2} \, (3 + n_s)$
(since the matter power spectrum $\Delta_\delta \propto k^4 \Delta_\zeta \propto k^{3 + n_s}$). Therefore, we get
\beq
\label{eq:modecoupling delta}
\fnl^{\rm eff} = - \frac{5}{12} \, (3 + n_s) = - \frac{5}{3} - \frac{5}{12} \, (n_s - 1) = - \frac{5}{3} +\fnl \, .
\eeq
This is the squeezed limit mode coupling result found in the literature, e.g.~\cite{bartmatrio05}.

This leads to the following modulation of the small-scale variance on a given {\it coordinate} scale $R$,
\beq
\label{eq:coord modulation}
\left(\sigma^2_R\right)_{\varphi_l} = \left( 1 + \frac{12}{5} \, f_{\rm NL}^{\rm eff} \, \zeta_l \right) \, \sigma^2_R
= \left( 1  - 4 \, \zeta_l -  (n_s - 1) \, \zeta_l \right) \, \sigma^2_R \, .
\eeq

However, the halo density should depend on the small-scale perturbation statistics on a fixed {\it physical}
scale. The long-wavelength part of the spatial component of the metric is given by,
\bea
ds^2 &=& a^2(\tau) \, \left[ \left( 1 - 2 \hat{\psi}_l \right)\, d{\bf x}^2 + \partial_i \partial_j \chi_l \, dx^i dx^j\right] \nonumber \\
&=& a^2(\tau) \, \left[ \left( 1 + 2  \, \zeta_l \right)\, d{\bf x}^2 \right] \, ,
\eea
where in the second line we have used $\hat \psi_l = - \zeta_l$ under
the diffeomorphism defined in Eq.~\ref{eq:coord rescaling},
and we have again neglected gradients of the long mode.
Thus a physical distance is given in terms of coordinate distance in synchronous gauge by
\beq
a(\tau) \, \tilde{R} = a(\tau) \, \left( 1 + \, \zeta_l \right) \, R.
\eeq
This of course simply reflects the coordinate rescaling Eq.~(\ref{eq:coord rescaling}) we applied to generate
the long mode in the first place.
The small-scale variance on a fixed physical scale $a(\tau) \, \tilde{R}$ is now
\bea
\sigma^2_{\tilde{R} (1 - \, \zeta_l)} &=& \int d^3{\bf k} \, |W(k \, \tilde{R} \, (1 + 5/3 \, \varphi_l))|^2 \, P_\delta(k) \nonumber \\
&=& \int d^3{\bf \tilde{k}} \, |W(\tilde{k} \, \tilde{R} )|^2 \, P_\delta(\tilde{k} \, (1 + \zeta_l)) \nonumber \\
&=& \sigma^2_{\tilde{R}} \, \left[  1 + 3 \, \zeta_l \, + \, \zeta_l \, \frac{d\ln P_\delta({k})}{d\ln k}\right] \nonumber \\
&=& \sigma^2_{\tilde{R}} \, \left[  1 + 4 \, \zeta_l  + \, (n_s - 1) \, \zeta_l\right] \, .
\eea
Here, as in the above, we treat the power spectrum as a pure power law so that 
${d\ln P_\delta({k})}/{d\ln k} = n_s$, but is is trivial to extend this to power spectrum with running of the spectral index and
higher order corrections.

The variance on the right-hand-side above, $\sigma^2_{\tilde{R}}$, is evaluated on a fixed coordinate scale, but still gives the small-scale
variance at that scale in the synchronous gauge coordinate system. Eq.~(\ref{eq:coord modulation}) gives the modulation of
that variance by $\zeta_l$.
Putting it all together, we find for the variance on a fixed, physical scale,
and explicitly writing the dependence of the small-scale variance on $\zeta_l$,
\bea
\left(\sigma^2_{R = \tilde{R} (1 - \zeta_l)}\right)_{\varphi_l} &=& \left[  1  + 4 \, \zeta_l + (n_s - 1) \, \zeta_l \right] \, \left(\sigma^2_{\tilde{R}}\right)_{\varphi_l} \nonumber \\
&=& \left[   1  + 4 \, \zeta_l + (n_s - 1) \, \zeta_l \right] \, \left[  1  - 4 \, \zeta_l - (n_s - 1) \, \zeta_l \right] \, \sigma^2_{\tilde{R}} \nonumber \\
&=& \sigma^2_{\tilde{R}} \, .
\eea
Thus the variance has {\it no} dependence on the large-scale mode whatsoever.  As a consequence, there can be no scale-dependent bias in single-field inflation\footnote{The Newtonian limit of this result was also explained in~\cite{Kehagias:2013rpa}, in which case the coordinate transformation acts trivially.}.

We clearly see that this result is trivial. The transformation to
fixed physical scale simply corresponds to evaluating the small-scale variance in the frame where there is no large-scale mode,
i.e.~it is the variance of $\tilde{\delta}_s({\bf \tilde{x}})$ on a scale $\tilde{R}$, which by definition does not depend on $\varphi_l$.
Thus, we could have written down the result that $d\sigma^2_R/d\varphi_l = 0$ from the start. Instead, we had generated the mode coupling
in a less convenient coordinate system and then transformed it back to the original coordinate system.

Finally, we would like to again emphasize that these results are identical to those that we derive from perturbation theory in
synchronous gauge, see Appendix~\ref{app:pert}. 
Although the method used here (and in the next subsection) greatly simplify the calculation,
all approaches give precisely the same result as long as one is careful to define the halos in terms of a fixed {\it physical} scale.

\subsection{Scale-dependent bias from direct application of consistency conditions}
\label{subsec:direct}

A very useful observation emphasized in~\cite{Creminelli:2013mca} (see also~\cite{Fitzpatrick:2009ci}) is
that the procedure outlined above does not assume that the
original background is homogeneous.  Given a solution in a
specific gauge, we can find a new solution to the equations of motion
in the same gauge by performing a diffeomorphism.  This procedure is
linear in the long wavelength mode (and leading order in derivatives
of the long mode), but is non-perturbative in the short wavelength
modes.  We used this feature in the previous subsections to determine
the mixing between the long and short modes.  However, there is no
reason we cannot apply this procedure directly to the halo density
field without ever discussing the biasing relative to the matter
density.  The halos may be determined by some non-perturbative
behavior of the short modes, but to determine the influence of the
long mode, all we need to know is how the halo density transforms under
this diffeomorphism.

One can simply repeat the argument from the pervious section replacing
$\rho$ by $\rho_h $, the halo number density.
Working in synchronous gauge, we again find the long mode via
${\bf x} = \tilde{\bf x} \, (1-\zeta_l)$.  The halo density transforms under a spatial diff as
\beq
\bar{\rho}_h(\tau) \, \left(1 + \delta_h({\bf x}, \tau) \right) = \rho({\bf x}, \tau) = \tilde{\rho}({\bf \tilde{x}}, \tau)
=  \bar{\rho}_h(\tau) \, \left( 1 + \tilde{\delta}_h({\bf \tilde{x}},  \tau) \right), 
\eeq
so that $\delta_s$ transforms as a scalar (as did $\zeta_s$) under the dilation,
\beq
\delta_h({\bf x},\tau) = \tilde{\delta}_h({\bf \tilde{x}}, \tau) \, .
\eeq
Thus, in the coordinate system with the long mode explicit we find
\beq
\delta_h({\bf x},\tau) = \tilde{\delta}_h({\bf x} (1 + \zeta_l)) = \tilde{\delta}_h({\bf x}) + \zeta_l \, {\bf x} \cdot \nabla \tilde{\delta}_h({\bf x}) =\tilde{\delta}_h({\bf x}) +{\cal O}(\delta_h \delta) .
\eeq
Notice that the contribution from the long mode only enters at quadratic order in the fluctuations.
Therefore we can conclude at linear order that 
\beq
\boxed{ \delta_h^{\rm sync.} =\tilde \delta_{h, S} ^{\rm sync}  \, \, \to \, \, \bng(k) = 0 }
\eeq
As discussed in Appendix~\ref{app:gauge}, in other gauges (e.g. Newtonian gauge) there may contributions at linear order that are proportional to $\partial_\tau\bar \rho_h$.  However, these terms appear for densities at fixed coordinate time, rather than proper time.  These linear contributions will cancel out when expressed in terms of observable quantities which are gauge invariant, as we explained in Section~\ref{sec:recipe}.

\section{Summary and Discussion}\label{sec:disc}

Scale-dependent bias in the clustering of galaxies or halos is a powerful probe
of primordial non-Gaussianity and therefore of the physics of inflation.
Specifically, constraining local non-Gaussianity with a precision $\sigma(\fnl) \lesssim 1$
is, in principle, achievable with future surveys and
may distinguish between single-field and multi-field inflation models.
In order to correctly use such a measurement
to test single-field inflation,
an exact theoretical prescription for the observed galaxy overdensity, $\Delta_g$,
is needed. This prescription must take into account the fact that $\Delta_g$
is measured in terms of {\it observed} redshifts and {\it observed} angular positions of galaxies,
should be free of gauge ambiguities, and should include an exact prediction
for the level of scale-dependent bias in single-field inflation.
The latter question in particular
has been a source of confusion and the main contribution of the present paper
has been to clarify the calculation of this scale-dependent bias
and to derive a consistent answer in several independent ways.
The final result, presented in Section \ref{sec:bias},
is that there is {\it zero} scale-dependent bias, $\bng \equiv 0$,
if inflation is governed by a single field, leading to the complete prescription
for galaxy clustering summarized in Section \ref{sec:recipe}.

Most of the confusion about scale-dependent bias in single-field models
stems from the fact that matter perturbations
in, for example, synchronous gauge have
order-unity mode coupling,
$\fnl^{\rm eff} = -5/3 - 5/12 \, (n_s - 1)$.
Since scale-dependent bias arises from a modulation of 
the initial small-scale variance
of matter perturbations ($\sigma_R^2$) by a large-scale curvature perturbation ($\zeta_l$),
at first sight the mode coupling would thus appear to produce a scale-dependent bias
corresponding to $\fnl$ of order unity.

However, we have argued that this is not the correct interpretation.
A crucial realization is that it can only be the initial variance of matter perturbations
on a given {\it physical} scale, $\tilde{R} = (1 + \zeta_l) \, R$,
that determines the halo number density
in that region at late times. While the initial variance of matter perturbations on a
fixed {\it coordinate}
scale does indeed see the order-unity modulation mentioned above,
we have shown explicitly in Section \ref{subsec:modecoupling} that
when the variance is converted from being evaluated at fixed coordinate scale $R$
to being evaluated at
fixed physical scale $\tilde{R}$,
the terms depending on $\zeta_l$ cancel and the modulation
is identically zero (with even the term proportional to $n_s - 1$ vanishing).

To obtain the above result, we have derived the synchronous gauge mode coupling of matter perturbations
in two ways. In Appendix \ref{app:pert}, we followed the approach common in the literature,
which matches the second order perturbation theory solution in the post-inflationary epoch to the
inflationary consistency relation.
However, in Section \ref{subsec:modecoupling}, we derived the same result
in a simpler and, we argue, more insightful way.
In single-field inflation, the local effect of a large-scale curvature perturbation $\zeta_l$
(i.e.~up to gradients)
on the physics on small scales
is equivalent to a simple coordinate transformation.
In other words, there is no local physical effect.
This follows from the fact that inflation is described by a single clock
in the single-field case and each region of space finds itself on the same attractor solution.
It remains true after inflation as long as perturbations evolve adiabatically,
which is always true on super-horizon scales (more generally, it is true on scales above
the sound horizon).
Using this principle, which is equivalent to the inflationary consistency
condition(s), we readily generated the well known mode coupling of synchronous gauge matter perturbations by
a diffeomorphism,
$\fnl^{\rm eff} = -5/12 \, d\ln\Delta^2_\delta(k)/d\ln k= -5/3 - 5/12 \, (n_s - 1)$ (see the derivation of Eq.~(\ref{eq:modecoupling delta})).
In this picture, the aforementioned cancellation of the $\zeta_l$ dependence
of the variance
became trivial: when converting to the initial variance of matter perturbations
on a fixed physical scale, we were simply undoing the diffeomorphism
that produced the synchronous gauge mode coupling in the first place.

Finally, in Section \ref{subsec:direct},
we derived the absence of scale-dependent bias
in an even simpler way by applying the diffeomorphism
generating $\zeta_l$ to the halo number density directly.
We demonstrated that, modulo gradients, the long mode does not affect the halo number
density at first order in perturbations and thus cannot generate a scale-dependent bias.

We would like to emphasize that the
absence of scale-dependent bias in single-field inflation has been to various degrees
understood in the vast
literature on consistency conditions in large-scale structure (e.g.~\cite{Peloso:2013zw, Kehagias:2013yd, Creminelli:2013mca,Peloso:2013spa,Creminelli:2013poa,Kehagias:2013rpa,Nishimichi:2014jna,Horn:2014rta,
  Ben-Dayan:2014hsa,Mirbabayi:2014gda, Horn:2015dra, Kehagias:2015tda,
  Nishimichi:2015kla, Dai:2015jaa}).
It was for example appreciated by many authors that the effect of the long mode on various quantities is zero when
expressed in terms of physical scales~\cite{Senatore:2012wy}, in convenient coordinates like Fermi-normal
coordinates~\cite{Baldauf:2011bh, Pajer:2013ana,Dai:2015jaa} or in Lagrangian space~\cite{Horn:2015dra}.  
The main contribution of the present paper was to resolve the paradox presented by the
order unity mode coupling in coordinate space that led others to conclude there must be scale-dependent bias of order unity even
in the single-field context. 
Moreover, we provided an explicit recipe for how to use this result to compute the {\it observed}
galaxy perturbations taking into account relativistic terms (see also~\cite{Kehagias:2015tda}).  The simple recipe we provide can be applied in any gauge and can therefore be easily incorporated into any existing calculation.  
We thus now have the tools to properly interpret
observational constraints on scale-dependent bias even when those constraints reach a precision
$\sigma(f_{\rm NL}) \lesssim 1$. If such constraints favor non-zero scale-dependent bias
as carefully defined in this paper, it would rule out the class of single-field inflation (modulo
the caveats that we have assumed the inflationary evolution to have been on the attractor solution).

\subsubsection*{Acknowledgements}

We thank the participants of the workshop {\it Testing Inflation with
Large Scale Structure: Connecting Hopes with Reality} held at CITA,
University of Toronto in October 2015 for stimulating discussions. We
particularly thank Matias Zaldarriaga for several insights and for
highlighting  the relevance of proper physical scales in this problem. Part of the
research described in this paper was carried out at the  Jet Propulsion
Laboratory, California Institute of Technology,  under a contract with
the National Aeronautics and Space  Administration. This work is
supported by NASA ATP grant 11-ATP-090. D.G. is supported by a NSERC
Discovery Grant. 

\newpage

\appendix

\section{Consistency Conditions in Newtonian and Synchronous Gauge}\label{app:gauge}

In this appendix, we will review how the adiabatic modes arise from diffeomorphisms that leave Newtonian gauge fixed.
We will then repeat the analysis in synchronous gauge and discuss the technical differences.  Finally we will compare the description of halos in each gauge and their relation to physical observables.

To derive the consistency conditions in conformal Newtonian gauge, we will follow~\cite{Creminelli:2013mca} and define the metric to take the form
\beq
ds^2 = a^2(\tilde \tau) [ -(1+2 \tilde \Phi) d\tau^2 + (1-2\tilde \Psi) \delta_{ij} d\tilde x^i d\tilde x^j] \ .
\eeq
Now let us consider the transformation
\bea
 \tau &=& \tilde \tau+ \epsilon(\tilde \tau)\\
 \x &=&\tilde  \x (1- \lambda)
\eea
where $\epsilon(\tau)$ is an arbitrary function of time and $\lambda$ is a constant.  This transformation shifts the potentials
\bea
 \Phi(\x,\tau) &=&\tilde \Phi(\x,\tau)  - \epsilon' - \H \epsilon \\
 \Psi(\x, \tau) &=& \tilde \Psi(\x,\tau)  - \lambda + \H \epsilon  \ ,
\eea
but leaves the gauge fixed.  Under this change of coordinates, any scalar, $s$, will transform (to linear order) as
\beq
 s^{\rm N} (\tau,  \x)= \tilde s(\tau,  \x) - \epsilon(\tau) \, \tilde s'(\tau,\x)  \ ,
\eeq
where the $^{\rm N}$ is a reminder that this is the scalar fluctuation in Newtonian gauge and $' \equiv \partial_\tau$.  If we start from a solution to Einstein's equations, any such transformation will generate a new solution.  However, most such solutions cannot be extended to physical solutions with large, but finite wavelengths $\{ \epsilon(\tau), \lambda \} \to\{ \epsilon(\tau, x), \lambda(x) \}$.  The reason is that some of the equations of motion may simply vanish when $k = 0$ and are not satisfied for any $k\neq 0$.  In conformal Newtonian gauge, the only such equation is $k^2(\Phi -\Psi)=0$.  Therefore, if we impose $\Phi = \Psi$ as an additional constraint, this transformation can be extended to a physical solution.  This requires that $\epsilon' + 2 \H \epsilon = \lambda$ or
\beq
\epsilon = \frac{\lambda}{a^2} \int^\tau d\tau' a^2(\tau') \equiv D_v  \, \lambda \  ,
\eeq
where $D_v$ is velocity growth function.  In terms of $\zeta = - \Psi - H \delta \rho / \dot \rho$ this means that $\zeta = \tilde \zeta + \lambda$ (where we used $\delta \rho=- \rho' \epsilon$) and therefore we see that from a differmorphism we have generated a long wavelength $\zeta$.  Therefore, if we start from a FRW solution and make the identification $\lambda = \zeta_l$ and $\tilde \zeta= 0$,  we generate a new solution
\bea
\Phi &=& - \zeta_l \, ( D_v' + \H D_v) \\
\Psi &=& - \zeta_l \, (1- \H D_v)   \\
\delta s^{\rm N} &=& - \zeta_l \, D_v  s' \ .
\eea
More generally, new solutions that are linear in $\zeta_l$ can be generated from any solution for the short modes, even non-perturbative solutions.  This procedure can also be extended to include a gradient of the long mode~\cite{Creminelli:2013mca} (via the conformal consistency conditions~\cite{Creminelli:2012ed,Hinterbichler:2012nm}) and to include contributions at non-linear order in the long mode~\cite{Mirbabayi:2014zpa,Joyce:2014aqa}.  We can then use this procedure to determine the mode coupling between long and short modes provided only that we neglect contributions ${\cal O}(\partial_i \partial_j \zeta_l)$.

Although the details of the above argument were specific to Newtonian gauge, such an argument will work in any gauge.  For our purposes, we are are interested in  synchronous gauge where the metric is taken to be
\bea
ds^2 &=& a^2(\tau) \left[ -  d\tau^2 + (1 - 2 \hat{\psi}) \, \delta_{ij}^{\rm K} \, dx^i dx^j +  \partial_i \partial_j \chi \, dx^i dx^j\right] 
\eea
It is easy to check that the transformation
\beq
 \x =\tilde \x (1- \zeta_l) 
\eeq
leaves the gauge unchanged and shifts the potential
\beq
\hat{\psi} \to \hat{\psi} - \zeta_l \ .
\eeq
Unlike Newtonian gauge, there is no change to the time coordinate and therefore the transformation law for a scalar is simply
\beq
\delta s^{\rm S} = 0 \ ,
\eeq
where $^{\rm S}$ is a reminder that this is the fluctuation in synchronous gauge.  A priori, one may have been concerned that this was not a physical solution, as much of the gauge freedom has not been fixed by the choice of metric.  Nevertheless, we can again check that in terms of $\zeta =-\hat \psi - H \delta \rho / \dot \rho$ we again have $\zeta \to \zeta + \zeta_l$.  This shows that the usual redundancies of synchronous gauge are playing no role in this transformation as those redundancies leave $\zeta$ fixed (after all, $\zeta$ is often called ``gauge invariant" because it does not transform under the redundancies of synchronous gauge).  One added benefit of this gauge redundancy in synchronous gauge is that we do not need to impose an extra condition on our transformation in order to extend it to a physical solution (one can check that there are no equations of motion that vanish when $k=0$).    

A priori, one might think that to see theses results in different gauges, we may simply look up the diffeomorphism that takes us from one gauge to another.  However, since our solution is equivalent to a diffeomorphism that leaves the gauge fixed, the diffeomorphism that takes us between gauges need not preserve the form of the long wavelength mode.  As a result, to understand the implications of the long mode in a given gauge, it is most useful to work directly in the appropriate gauge from the beginning and add the long mode using the relevant diffeomorphism that keeps that gauge fixed.  

\vskip 8pt
Now let us consider the implications for the matter density contrast $\delta$ in synchronous gauge.  The total matter density $\rho(x)$ is a scalar under diffeomorphisms and therefore
\beq
 \rho^{\rm S}( \x,t )  =  \tilde \rho_S^{\rm S}(\x, t) + \zeta_l(\x) \x \cdot {\bf \partial} \tilde \rho^{\rm S}_s + \ldots   \ .
\eeq
We see that we have generated mode coupling between the short mode, $\rho_s$, and the long wavelength mode, $\zeta_l$, as described in the main text.  We can perform a similar calculation in Newtonian gauge where we find 
\beq
 \rho^{\rm N}( \x, \tau )  = \tilde \rho^{\rm N}_s(\x, \tau ) +\zeta_l  D_v  \tilde \rho'+ \zeta_l \, \x \cdot {\bf \partial} \tilde \rho^{\rm N}_s \ .
\eeq
Defining $\delta^{\rm S, N} = \frac{\rho^{\rm S, N}}{\bar \rho} -1$ where $\bar \rho$ is the homogenous solution, we find
\beq
\delta^{\rm S} = \delta_s^{\rm S} + \zeta_L(\x) \x \cdot {\bf \partial} \delta^{\rm S}_s \qquad \delta^{\rm N} = \delta_S^{\rm N} + D_v \zeta_l ( \delta' + \frac{\bar\rho'}{\bar \rho}(1+\delta_s) )  + \zeta_l \x \cdot {\bf \partial} \delta^{\rm N}_s
\eeq
In both gauges, we find a coupling between the long and short modes.  However, in Newtonian gauge we find additional linear terms in $\zeta_l$ as well as extra mode coupling contributions which arise from the time component of the diffeormorphism.

From the above procedure is should be clear that nothing required that $\rho$ was the dark matter density.  It could have been any scalar quantity, including the density of halos or galaxies.  Therefore, without doing any additional work, we see that at linear order in $\delta_g$ and $\zeta_l$ we have 
\beq
\delta_g^{\rm S} = \delta_{h, s} ^{\rm S} + {\cal O}(\delta_g \zeta_l) \, \, \to \, \, \bng = 0 
\eeq
where $\delta_h^{\rm S}$ is the physical density contrast.  We emphasize that this means that, in the presence of a long wavelength mode, there is no change to density contrast of halos in synchronous gauge.  

One might worry that in Newtonian gauge,
we do find terms linear in $\zeta_l$ that would appear as scale dependent bias.
However, this feature arises because we are describing the density
on constant coordinate time slices, rather than in terms of some physical definition of time.
If we were to compute the observed galaxy density in a redshift survey,
one finds that the results appear in terms of the the gauge invariance density contrast
\beq
\delta_{\rm invariant} \equiv \delta_g + \frac{d\ln\bar{n}_g}{d\ln a} \, a H \delta \tau =  \delta^{\rm S}_g \ ,
\eeq
where the $\delta \tau$ arises from the fluctuations in the observed redshift relative to the coordinate time and therefore the last equality follows from the feature that $\delta \tau^{\rm S} \equiv 0$.  It is clear that there will be no scale dependent bias when written in terms of observable quantities in any gauge, but this is most transparent in synchronous gauge.  

\section{Second Order Perturbation Theory Calculation of Mode Coupling}\label{app:pert}

In Section \ref{subsec:modecoupling}, we used a diffeomorphism to directly generate
the mode coupling of matter perturbations in synchronous gauge, $f_{\rm NL}^{\rm eff} = -5/3 - 5/12 \, (n_s - 1)$.
This result has been derived in the literature many times by solving the second order perturbation theory equations
for the matter perturbations in the post-inflationary Universe and by then matching this to the inflationary consistency condition
on $\zeta$. Here, we briefly review this alternative derivation in the simple case of a matter-only Universe
and show that it indeed leads to the same result, albeit through a more cumbersome calculation.
We will not rederive the second order perturbation theory solution itself from scratch, but instead use well known expressions from
the literature, specifically
from \cite{matmolbrun98,bartmatrio05}
(see also, e.g., \cite{bartetal10,brunietal14a}).

Let us expand the perturbations to second order as $\delta = \delta^{(1)} + 1/2 \, \delta^{(2)}$,
$\chi = \chi^{(1)} + 1/2 \, \chi^{(2)}$, $\hat{\psi} = \hat{\psi}^{(1)} + 1/2 \, \hat{\psi}^{(2)}$.
In synchronous gauge, in a matter-only universe,
the first order solution is then given by
\bea
\label{eq:1st order}
\chi^{(1)} &=& -\frac{\tau^2}{3} \, \varphi \nonumber \\
\hat{\psi}^{(1)} &=& \frac{5}{3} \, \varphi \nonumber \\
\delta^{(1)} &=& \frac{\tau^2}{6} \, \nabla^2 \varphi
\eea
(note that we have fixed the remaining gauge freedom present in synchronous gauge).

We split the solution into small- and large-scale components,
$\varphi = \varphi_s + \varphi_l$.
Then, to compute the mode coupling of interest, we need the second order part of the small-scale solution,
but only those terms that are products of one small-scale mode and one large-scale mode,
and only to zeroth order in gradients of $\varphi_l$.
Specifically, we care about terms of order $\varphi_l \, \delta_s^{(1)} \sim \varphi_l \tau^2 \nabla^2 \varphi_s$
contributing to $\delta_s^{(2)}$ and terms of order $\varphi_l \, \varphi_s$ contributing to $\hat{\psi}^{(2)}$.
The full calculation of the second order solution is cumbersome, if conceptually straightforward, and has been
done for us (see, e.g., Eqs.~(6) and (7) of \cite{bartmatrio05}). Focusing on the relevant terms, the general result
is the class of solutions\footnote{Note that \cite{matmolbrun98} appears to have a typo causing an inconsistency
between the solution for the metric perturbation and the matter overdensity.}
\beq
\label{eq:2pt gen}
\delta^{(2)} = \frac{20}{9} \, \varphi_l \, \tau^2 \, \nabla^2 \varphi_s + \frac{1}{10} \, \tau^2 \, \nabla^2 \hat{\psi}^{(2)}.
\eeq
Note that, to the order of interest, simply solving the second order perturbation equations during matter domination
allows us the freedom to choose
$\hat{\psi}^{(2)}$ (at least the relevant $\mathcal{O}(\varphi_l \, \varphi_s)$ contribution),
which is why $\delta^{(2)}$ is expressed in terms of it.
For example, the equations would be solved by
\bea
\label{eq:one sol}
\delta_s^{(2)} &=& \frac{20}{9} \, \varphi_l \, \tau^2 \, \nabla^2 \varphi_s  \\
\hat{\psi}_s^{(2)} &=& 0 \nonumber.
\eea
However, we can always apply a redefinition,
\beq
\varphi_s \equiv (1 +  \varphi_l \, \hat{a}) \, \tilde{\varphi}_s,
\eeq
express Eq.~(\ref{eq:one sol}) in terms of $\tilde{\varphi}_s$ and finally drop the tilde, to get a new second order solution,
\bea
\delta_s^{(2)} &=& \frac{20}{9} \, \varphi_l \, \tau^2 \, \nabla^2 \varphi_s +  \frac{\tau^2}{3} \, \varphi_l \, \nabla^2(\hat{a} \varphi_s) \nonumber \\
\hat{\psi}^{(2)}_s &=& \frac{10}{3} \, \varphi_l \,  \hat{a} \, \varphi_s \nonumber.
\eea
Thus, indeed $\delta^{(2)} = \frac{20}{9} \, \varphi_l \, \tau^2 \, \nabla^2 \varphi_s + \frac{1}{10} \, \tau^2 \, \nabla^2 \hat{\psi}^{(2)}$.

While the matter domination second-order perturbation theory calculation thus does not uniquely specify the solution,
we can fix it by requiring that the first order solution is Gaussian, which, in terms of mode coupling, means
that $\varphi_s^{(1)}$ is statistically independent of $\varphi_l^{(1)}$ (note that this is already done in
the solution Eq.~(6) and (7) in \cite{bartmatrio05}). We do this by matching the solution
to the consistency relation for $\zeta$ (see Section \ref{subsec:maldacena}).
Recall that the comoving curvature perturbation on constant density slices is defined as
\beq
e^{2 \zeta} = \left(1 - 2 \hat{\psi}\right)_{\delta \rho = 0}.
\eeq
We match the perturbation theory solution Eq.~(\ref{eq:2pt gen}) to $\zeta$, in the regime where both the large and small modes are far outside the horizon.
We can then ignore gradients acting on both $\varphi_l$ and $\varphi_s$.
In this limit,
$\hat{\psi}^{(2)}|_{\delta \rho = 0} = \hat{\psi}^{(2)}$, the right-hand side being the solution in synchronous gauge (this follows from, e.g., Eqs.~(131) and (134) of \cite{bartetal04}).
Thus,
\beq
\hat{\psi} = -\zeta - \zeta^2,
\eeq
or in terms of the mode coupling,
\beq
\hat{\psi}_s = -\zeta_s \, (1 + 2 \zeta_l).
\eeq
The consistency relation for $\zeta$ says
\beq
\zeta_s = \left(1 +  \zeta_l \, {\bf x} \cdot \nabla \right)\, \zeta_s^G.
\eeq
Therefore,
\beq
\hat{\psi}_s = - \left(1 + 2 \zeta_l + \zeta_l \, {\bf x} \cdot \nabla\right) \, \zeta_s^G,
\eeq
where $\zeta_s^G$ is independent of $\zeta_l$ so that this gives the true statistical mode coupling.
Identifying the first order perturbation theory solution with the Gaussian field, i.e.~$\psi^{(1)} = \frac{5}{3} \, \varphi \equiv - \zeta^G$,
now fixes the second order solution to
\bea
\label{eq:2nd sol}
\hat{\psi}^{(2)} &=& - \frac{100}{9} \, \varphi_l \, \left( 1 + \frac{1}{2} \, {\bf x} \cdot {\nabla} \right) \, \varphi_s \nonumber \\
\delta^{(2)}_s &=& \frac{20}{9} \, \varphi_l \, \tau^2 \, \nabla^2 \varphi_s - \frac{10}{9}\, \varphi_l \, \tau^2 \, \nabla^2\left[\left(1 + \frac{1}{2}\, {\bf x} \cdot {\nabla} \right) \varphi_s \right] \nonumber \\
&=& \frac{20}{9} \, \varphi_l \, \tau^2 \, \nabla^2 \varphi_s - \frac{10}{9}\, \varphi_l \, \tau^2 \, \left(2 + \frac{1}{2}\, {\bf x} \cdot {\nabla} \right) \, \nabla^2 \varphi_s \nonumber \\
&=&  -\frac{5}{9} \, \varphi_l \, \tau^2 \, {\bf x} \cdot {\nabla}(\nabla^2 \varphi_s) \nonumber \\
&=& -\frac{10}{3} \, \varphi_l \, \tau^2 \, {\bf x} \cdot {\nabla} \,\delta_s^{(1)} = 2 \, \zeta_l \, {\bf x} \cdot {\nabla} \,\delta_s^{(1)},
\eea
with $\delta_s^{(1)}$ independent of the long mode by definition.  

We see that this is precisely the same result for the mode coupling that we derived directly using the consistency conditions,
Eq.~(\ref{eq:diff mode coupling}). As explained in Section \ref{subsec:modecoupling}, this is equivalent to local
non-Gaussianity with $f_{\rm NL}^{\rm eff} = -5/3 - 5/12 \, (n_s - 1)$.
However, as detailed in Section \ref{subsec:modecoupling}, the mode coupling disappears when the short wavelength modes
are evaluated on a fixed physical scale as opposed to a fixed (synchronous gauge) coordinate scale.

\newpage
\addcontentsline{toc}{section}{References}
\bibliographystyle{utphys}
\bibliography{biasrefs}
\end{document}